\font\bbb=msbm10                                                   

\def\C{\hbox{\bbb C}}
\def\N{\hbox{\bbb N}}
\def\R{\hbox{\bbb R}}
\def\Z{\hbox{\bbb Z}}

\def\APL{{\sl Appl.\ Phys.\ Lett.}}
\def\APS{{\sl Acta Physica Slovaca}}

\def\CS{{\sl Complex Systems}}

\def\EL{{\sl Europhys.\ Lett.}}

\def\IJMPC{{\sl Int.\ J. Mod.\ Phys.\ C}}
\def\IJTP{{\sl Int.\ J. Theor.\ Phys.}}

\def\JPA{{\sl J. Phys.\ A:  Math.\ Gen.}}

\def\JSP{{\sl J. Stat.\ Phys.}}
\def\LAS{{\sl Los Alamos Science}}

\def\NPB{{\sl Nucl.\ Phys.\ B}}

\def\PD{{\sl Physica D}}

\def\PRA{{\sl Phys.\ Rev.\ A}}

\def\PRE{{\sl Phys.\ Rev.\ E}}
\def\PRL{{\sl Phys.\ Rev.\ Lett.}}

\def\Sc{{\sl Science}}
\def\SPJETP{{\sl Sov.\ Phys.\ JETP}}

\def\TMP{{\sl Theor.\ and Math.\ Phys.}}

\def\ZP{{\sl Z. Physik}}

\def\dajm{\hbox{D. A. Meyer}}

\def\brosl{\hbox{B. Hasslacher}}

\def\baxter{\hbox{R. J. Baxter}}
\def\feynman{\hbox{R. P. Feynman}}

\def\hfb{\hfil\break}

\catcode`@=11
\newskip\ttglue

   \font\ninerm=cmr9    \font\eightrm=cmr8   \font\sixrm=cmr6
  \font\ninebf=cmbx9   \font\eightbf=cmbx8  \font\sixbf=cmbx6
\font\tenit=cmti10  \font\nineit=cmti9   \font\eightit=cmti8  
  \font\ninesl=cmsl9   \font\eightsl=cmsl8  
  \font\ninemi=cmmi9   \font\eightmi=cmmi8  \font\sixmi=cmmi6

\font\bigtenbf=cmr10 scaled\magstep2 

\def\ninepoint{\def\rm{\fam0\ninerm}%
  \textfont0=\ninerm \scriptfont0=\sixrm
  \textfont1=\ninemi \scriptfont1=\sixmi
  \textfont\itfam=\nineit  \def\it{\fam\itfam\nineit}%
  \textfont\slfam=\ninesl  \def\sl{\fam\slfam\ninesl}%
  \textfont\bffam=\ninebf  \scriptfont\bffam=\sixbf
    \def\bf{\fam\bffam\ninebf}%
  \tt \ttglue=.5em plus.25em minus.15em
  \normalbaselineskip=11pt
  \setbox\strutbox=\hbox{\vrule height8pt depth3pt width0pt}%
  \normalbaselines\rm}

\def\eightpoint{\def\rm{\fam0\eightrm}%
  \textfont0=\eightrm \scriptfont0=\sixrm
  \textfont1=\eightmi \scriptfont1=\sixmi
  \textfont\itfam=\eightit  \def\it{\fam\itfam\eightit}%
  \textfont\slfam=\eightsl  \def\sl{\fam\slfam\eightsl}%
  \textfont\bffam=\eightbf  \scriptfont\bffam=\sixbf
    \def\bf{\fam\bffam\eightbf}%
  \tt \ttglue=.5em plus.25em minus.15em
  \normalbaselineskip=9pt
  \setbox\strutbox=\hbox{\vrule height7pt depth2pt width0pt}%
  \normalbaselines\rm}

\def\sfootnote#1{\edef\@sf{\spacefactor\the\spacefactor}#1\@sf
      \insert\footins\bgroup\eightpoint
      \interlinepenalty100 \let\par=\endgraf
        \leftskip=0pt \rightskip=0pt
        \splittopskip=10pt plus 1pt minus 1pt \floatingpenalty=20000
        \parskip=0pt\smallskip\item{#1}\bgroup\strut\aftergroup\@foot\let\next}
\skip\footins=12pt plus 2pt minus 2pt
\dimen\footins=30pc

\def\ie{{\it i.e.}}
\def\eg{{\it e.g.}}

\def\etal{{\it et al.}}

\def\Shor{1}
\def\reviews{2}
\def\Schumacher{3}
\def\gates{4}
\def\iontraps{5}
\def\cavityQED{6}
\def\NMRexp{7}
\def\qdots{8}
\def\MargolusCAM{9}
\def\QCA{10}
\def\qcaqlg{11}
\def\Watrous{12}
\def\MargolusBB{13}
\def\LGAsim{14}
\def\qmI{15}
\def\BoghosianTaylorsim{16}
\def\BoghosianTaylorSeq{17}
\def\Feynman{18}
\def\nanoarch{19}
\def\Biafore{20}
\def\lgbu{21}
\def\DestrideVega{22}
\def\Bethe{23}
\def\Baxter{24}
\def\boundaryR{25}
\def\Riazanov{26}
\def\VannesteSebbahSornette{27}
\def\SquierSteiglitz{28}
\def\BenjaminJohnson{29}

\magnification=1200
\input epsf.tex

\dimen0=\hsize \divide\dimen0 by 13 \dimendef\chasm=0
\dimen1=\hsize \advance\dimen1 by -\chasm \dimendef\usewidth=1
\dimen2=\usewidth \divide\dimen2 by 2 \dimendef\halfwidth=2
\dimen3=\usewidth \divide\dimen3 by 3 \dimendef\thirdwidth=3
\dimen4=\hsize \advance\dimen4 by -\halfwidth \dimendef\secondstart=4
\dimen5=\halfwidth \advance\dimen5 by -10pt \dimendef\indenthalfwidth=5
\dimen6=\thirdwidth \multiply\dimen6 by 2 \dimendef\twothirdswidth=6
\dimen7=\twothirdswidth \divide\dimen7 by 4 \dimendef\qttw=7
\dimen8=\qttw \divide\dimen8 by 4 \dimendef\qqttw=8
\dimen9=\qqttw \divide\dimen9 by 4 \dimendef\qqqttw=9

\parskip=0pt\parindent=0pt

\line{\hfil October 1996}
\line{\hfil{\it revised\/} November 1997}
\line{\hfil quant-ph/9712052}
\bigskip\bigskip
\centerline{\bf\bigtenbf QUANTUM MECHANICS OF LATTICE GAS AUTOMATA}
\bigskip
\centerline{\bf\bigtenbf II.  BOUNDARY CONDITIONS}
\bigskip
\centerline{\bf\bigtenbf AND OTHER INHOMOGENEITIES}
\vfill
\centerline{\bf David A. Meyer}
\bigskip 
\centerline{\sl Institute for Physical Sciences}
\smallskip
\centerline{\sl and}
\smallskip
\centerline{\sl Project in Geometry and Physics}
\centerline{\sl Department of Mathematics}
\centerline{\sl University of California/San Diego}
\centerline{\sl La Jolla, CA 92093-0112}
\centerline{dmeyer@chonji.ucsd.edu}
\vfill
\centerline{ABSTRACT}
\bigskip
\noindent We continue our analysis of the physics of quantum lattice
gas automata (QLGA).  Previous work has been restricted to periodic or
infinite lattices; simulation of more realistic physical situations 
requires finite sizes and non-periodic boundary conditions.  
Furthermore, envisioning a QLGA as a nanoscale computer architecture 
motivates consideration of inhomogeneities in the `substrate'; this 
translates into inhomogeneities in the local evolution rules.  
Concentrating on the one particle sector of the model, we determine
the various boundary conditions and rule inhomogeneities which are 
consistent with unitary global evolution.  We analyze the reflection 
of plane waves from boundaries, simulate wave packet refraction across 
inhomogeneities, and conclude by discussing the extension of these 
results to multiple particles.
\bigskip
\global\setbox1=\hbox{PACS numbers:\enspace}
\global\setbox2=\hbox{PACS numbers:}
\parindent=\wd1
\item{PACS numbers:}  03.65.-w,  
                      02.70.-c,  
                      11.55.Fv,  
                      89.80.+h.  
\item{\hbox to \wd2{KEY\hfill WORDS:}}   
                      quantum lattice gas; quantum cellular automaton; 
                      quantum computation; 
\item{}               boundary conditions; inhomogeneities.

\vfill
\eject

\headline{\ninepoint\it Quantum mechanics of LGA \hfil David A. Meyer}
\parskip=10pt
\parindent=20pt

\noindent{\bf 1.  Introduction}

\noindent Shor's discovery of a polynomial time quantum algorithm for
factoring [\Shor] stimulated a surge of interest in quantum 
computation (see the extensive bibliographies of [\reviews]).  Most 
work has concentrated on serial algorithms---sequences of unitary, few
qubit%
\sfootnote*{A {\sl qubit\/} [\Schumacher] is a quantum system whose 
            state is a vector in a two dimensional Hilbert space, \eg,
            a spin-${1\over2}$ particle fixed in space.}
operations---the quantum version of serial Boolean logic [\gates].  
Single quantum logic gates have been realized experimentally in ion 
traps [\iontraps] and quantum electrodynamics cavities [\cavityQED], 
and short sequences of such unitary operations have recently been 
implemented with NMR [\NMRexp].  All of these systems, as well as
proposed solid state architectures such as arrays of quantum dots
[\qdots], exist physically in $d > 0$ spatial dimensions and therefore
naturally evolve in {\sl parallel}.  Imposing a single gate operation
restricts the rest of the qubits to be invariant, \ie, they must 
evolve by the identity operator; at the opposite extreme all the 
qubits would evolve according to the same, local (few qubit) operation
during a single timestep.  A quantum computer evolving according to
such a homogeneous, local, unitary rule would have the quantum version
of the massively parallel architecture possessed, for example, by
Margolus' CAM machines [\MargolusCAM].

The simplest algorithms which would run on such an architecture are 
quantum cellular automata (QCA) [\QCA] or quantum lattice gas automata 
(QLGA) [\qcaqlg].  Even in $d = 1$ spatial dimensions QCA are capable 
of universal computation [\Watrous], and the existence of the 
universal reversible billiard ball computer [\MargolusBB] implies that
QLGA are also, at least in $d \ge 2$ spatial dimensions.  Just as
classical LGA are most effectively deployed to simulate physical 
systems such as fluid flow [\LGAsim], however, QLGA most naturally
simulate quantum physical systems [\qcaqlg,\qmI,\BoghosianTaylorsim]:
with the simplest homogeneous evolution rule, one particle QLGA 
simulate the constant potential Dirac [\qcaqlg] or Schr\"odinger 
[\BoghosianTaylorSeq] equation, depending on the relative scaling of
the lattice spacing and timestep.

An earlier paper [\qmI] initiated a project to analyze which physical 
processes QLGA can simulate effectively.  In that paper and in this 
one we concentrate on the most general model for a single quantum 
particle with speed no more than 1 in lattice units, moving on a 
lattice in one dimension.  The amplitudes for the particle to be 
(left, right) moving at a lattice point $x \in L$ are combined into a 
two component complex vector 
$\psi(t,x) := \bigl(\psi_{-1}(t,x),\psi_{+1}(t,x)\bigr)$ which evolves 
as
$$
\psi(t+1,x) 
= w_{-1}\psi(t,x-1) + w_0\psi(t,x) + w_{+1}\psi(t,x+1).     \eqno(1.1)
$$
Here the weights $w_i \in M_2(\C)$ are $2 \times 2$ complex matrices 
constrained by the requirement that the global evolution matrix
$$
U :=
\pmatrix{
\ddots&        &        &        &        &        &      \cr
      & w_{-1} & w_{ 0} & w_{+1} &        &        &      \cr
      &        & w_{-1} & w_{ 0} & w_{+1} &        &      \cr
      &        &        & w_{-1} & w_{ 0} & w_{+1} &      \cr
      &        &        &        &        &        &\ddots\cr
}                                                           \eqno(1.2)
$$
be unitary.  We showed in [\qcaqlg] that the most general parity 
invariant solution, up to unitary equivalence and an overall phase, is 
given by
$$
\setbox1=\hbox{%
$
w_{-1} = \cos\rho\pmatrix{0 & i\sin\theta \cr
                          0 &  \cos\theta \cr
                         }
\qquad
w_{+1} = \cos\rho\pmatrix{ \cos\theta & 0 \cr
                          i\sin\theta & 0 \cr
                         }
$
}
\eqalign{
w_{-1} = \cos\rho\pmatrix{0 & i\sin\theta \cr
                          0 &  \cos\theta \cr
                         }
\qquad
w_{+1} = \cos\rho\pmatrix{ \cos\theta & 0 \cr
                          i\sin\theta & 0 \cr
                         }                                         \cr
\hbox to\wd1{\hfil%
$
w_0  =   \sin\rho\pmatrix{  \sin\theta & -i\cos\theta \cr
                          -i\cos\theta &   \sin\theta \cr
                         }.
$ 
\hfil}                                                             \cr
}                                                           \eqno(1.3)
$$

Describing the evolution by (1.1)--(1.3) assumes that the system is
homogeneous in space and that the lattice $L$ is isomorphic either to 
the integers $\Z$ or to a periodic quotient thereof, say $\Z_N$.  To 
simulate physical systems [\Feynman] more generally, the model should be 
extended to allow for finite size and non-periodic boundary 
conditions.  Furthermore, envisioning a QLGA as a nanoscale quantum 
computer architecture [\nanoarch,\Biafore,\reviews] motivates consideration of 
inhomogeneities in the `substrate', possibly as a step towards 
implementing logical gates [\gates] and away from simply simulating quantum
physical systems.  In [\qmI] we showed how to introduce an 
inhomogeneous potential in the model; the purpose of this paper is to 
investigate more general inhomogeneities in the evolution rule, 
including boundary conditions.

In Section~2 we consider the simplest possible modification of the
evolution rule (1.1) for a boundary at $x = 0$, say, setting
$w_{-1}$ there to 0 and allowing the weight $\overline{w}_0$ to differ 
from the constant $w_0$ of the rest of the lattice.  The resulting 
Type~I boundary condition suggests the form for a corresponding Type~I 
inhomogeneity where the global evolution matrix (1.2) is changed by
replacing one of the $w_0$ blocks with a different matrix
$\widehat{w}_0$, and allowing the weights $w_i$ to differ on either 
side of the antidiagonal through it.  In Section~3 we show that such
an inhomogeneous rule is unitary provided $\theta$ is the same for all
the weights.

There is a dual inhomogeneity across which $\rho$ is constant but 
$\theta$ may differ; we describe this Type~II inhomogeneity in 
Section~4 and find the corresponding boundary condition.  In Section~5 
we observe that the Type~I and II inhomogeneities can occur together,
changing both $\rho$ and $\theta$.  The corresponding Type~III 
boundary condition has an extra degree of freedom, justifying distinct
classification.

In Section 6 we show how to find the eigenfunctions of $U$ in the
presence of these boundaries.  In each case the result is a linear 
combination of left and right moving plane waves with the same 
frequency.  On a finite lattice with two boundaries, the spectrum of
$U$ is discrete.  In Section 7 we investigate the discrete spectra for 
pairs of each type of boundary condition, determining how they depend 
on the boundary parameters and what the consequences are for the 
eigenfunctions.

Simulations of wave packets on lattices with boundary conditions and
in the presence of inhomogeneities confirms that the physical 
consequences of these inhomogeneous evolution rules are as expected.
We show some results in Section~8.  

We conclude in Section~9 with a summary and a discussion of the 
extension of this work to the multiple particle sector of the Hilbert
space.

\medskip
\noindent{\bf 2.  Type I boundary conditions}
\nobreak

\nobreak
\noindent If our system is neither infinite nor periodic, we must 
model it on a bounded lattice, \eg,
$L = \{ x \in \Z \mid 0 \le x \le N-1 \}$.  Since there is no lattice
point to the left of $0$, it is clear that the evolution rule (1.1)
must be adjusted there (as it must also be at the right boundary). 
Making the minimal change in the model, let us suppose that the global 
evolution matrix takes the form
$$
U :=
\pmatrix{
\overline{w}_{ 0} & w_{+1} &        &      \cr
           w_{-1} & w_{ 0} & w_{+1} &      \cr
                  & w_{-1} & w_{ 0} &      \cr
                  &        &        &\ddots\cr
},                                                          \eqno(2.1)
$$
where the $w_i$ are given by (1.3).  Thus a left moving particle at
$x = 1$ has the same amplitudes (given by $w_{+1}$) to advect to 
$x = 0$ and scatter to the left or right, and a right moving particle 
at $x = 0$ has the same amplitudes (given by $w_{-1}$) to advect to 
$x = 1$ and scatter to the left or right, as each would were there no 
boundary.  (The analogous form for the evolution rule at a right
boundary is obtained by a parity transformation.)  The only 
differences we allow for this {\sl Type~I\/} boundary condition are in 
the amplitudes for the evolution of a left moving particle at $x = 0$ 
and for the scattering of a right moving particle at $x = 0$ which 
remains there during the advection step; these are given by 
$\overline w_0$.

The unitarity conditions $UU^{\dagger} = I = U^{\dagger}U$ impose the
following constraints on $\overline{w}_0$:
$$
\eqalignno{
I &= \overline{w}_0^{\vphantom{\dagger}}
     \overline{w}_0^{\dagger} + 
     w_{+1}^{\vphantom{\dagger}} w_{+1}^{\dagger}           &(2.2a)\cr
0 &= \overline{w}_0^{\vphantom{\dagger}}
     w_{-1}^{\dagger} + 
     w_{+1}^{\vphantom{\dagger}} w_0^{\dagger}              &(2.2b)\cr
\noalign{\hbox{and}}
I &= \overline{w}_0^{\dagger}
     \overline{w}_0^{\vphantom{\dagger}} + 
     w_{-1}^{\dagger}w_{-1}^{\vphantom{\dagger}}            &(2.3a)\cr
0 &= w_{+1}^{\dagger}
     \overline{w}_0^{\vphantom{\dagger}} +
     w_0^{\dagger} w_{-1}^{\vphantom{\dagger}}.             &(2.3b)\cr
}
$$
Let
$$
\overline{w}_0 := \pmatrix{y_1 & y_2 \cr
                           y_3 & y_4 \cr
                          }.                                \eqno(2.4)
$$
Then, assuming $\cos\rho \not= 0$, $(2.2b)$ implies
$$
\eqalign{
y_2 &= -i\cos\theta \sin\rho                                       \cr
y_4 &=   \sin\theta \sin\rho,                                      \cr
}                                                           \eqno(2.5)
$$
while $(2.3b)$ implies 
$$
y_1 = i y_3 \tan\theta.                                     \eqno(2.6)
$$
The normalization condition $(2.2a)$ requires
$$
y_3 = -ie^{i\upsilon} \cos\theta                            \eqno(2.7)
$$
for some arbitrary phase angle $\upsilon \in \R$.  Combining 
(2.4)--(2.7), we find
$$
\overline{w}_0 
 = \pmatrix{  e^{i\upsilon} \sin\theta & -i\cos\theta \sin\rho \cr
            -ie^{i\upsilon} \cos\theta &   \sin\theta \sin\rho \cr
           },                                               \eqno(2.8)
$$
which satisfies all the constraints (2.2) and (2.3).

The Type~I boundary condition defined by (2.1) and (2.8) gives the 
same amplitudes as (1.3) for the scattering of a right moving particle 
at $x = 0$ which remains there; only the amplitudes for the scattering 
of a left moving particle at $x = 0$ differ from the no boundary 
situation.  The latter depend on a single real parameter $\upsilon$
characterizing the boundary.  Notice also that these amplitudes do not
vanish in the decoupled case $\rho = 0$ (whence $w_0 = 0$).  That is,
$\overline{w}_0 \not= 0$ is required to define unitary boundary 
conditions even when the particle has speed 1 everywhere else in the
lattice.

\medskip
\noindent{\bf 3.  Type I inhomogeneities}
\nobreak

\nobreak
\noindent The boundary weight $\overline{w}_0$ defined by (2.8) has
the same form as the weight $w_0$ defined in (1.3), except that the 
factor of $\sin\rho$ in the first column is replaced by 
$e^{i\upsilon}$.  Thus we can interpret the evolution rule defined by 
(2.1) and (2.8) as describing a system where the coupling constant
$\rho$ satisfies $\cos\rho = 0$ at and to the left of $x = 0$.  This 
would make $w_{-1} = 0 = w_{+1}$, so there would be no advection to 
the left of $x = 0$.  This suggests that the $\overline{w}_0$ we found 
in Section 2 may be a special case of an inhomogeneity in the coupling 
constant $\rho$.  So let us consider a {\sl Type~I\/} evolution rule 
inhomogeneity of the form:
$$
U :=
\pmatrix{
\ddots&         &         &         &         &         &       \cr
      & w_{-1}' & w_{ 0}' & w_{+1}' &         &         &       \cr
      &         & w_{-1}' & \widehat{w}_{0}^{\vphantom\prime}
                          & w_{+1}^{\vphantom\prime}    &         
                                                        &       \cr
      &         &         & w_{-1}^{\vphantom\prime}
                          & w_{ 0}^{\vphantom\prime}
                          & w_{+1}^{\vphantom\prime}    &       \cr
      &         &         &         &         &         &\ddots\cr
},                                                          \eqno(3.1)
$$
where the $w_i =: w_i(\rho,\theta)$ are defined by (1.3) and 
$w^{\prime}_i := w^{\vphantom\prime}_i(\rho',\theta')$.

Now the unitarity conditions impose constraints on the relation 
between the $w_i$ and the $w'_i$ as well as on the inhomogeneity 
matrix $\widehat{w}_0$:
$$
\eqalignno{
0 &= w_{+1}^{\prime\vphantom\dagger} w_{-1}^{\dagger}       &(3.2a)\cr
0 &= w_{-1}^{\prime\vphantom\dagger} w_0^{\prime\dagger} +
     \widehat{w}_0^{\vphantom\dagger}
     w_{+1}^{\prime\dagger}                                 &(3.2b)\cr
I &= w_{-1}^{\prime\vphantom\dagger} w_{-1}^{\prime\dagger} +
     \widehat{w}_0^{\vphantom{\dagger}}
     \widehat{w}_0^{\dagger} + 
     w_{+1}^{\vphantom{\dagger}} w_{+1}^{\dagger}           &(3.2c)\cr
0 &= \widehat{w}_0^{\vphantom{\dagger}}
     w_{-1}^{\dagger} + 
     w_{+1}^{\vphantom{\dagger}} w_0^{\dagger}              &(3.2d)\cr
\noalign{\hbox{and}}
0 &= w_{-1}^{\prime\dagger} w_{+1}                          &(3.3a)\cr
0 &= w_0^{\prime\dagger} w_{+1}^{\prime\vphantom\dagger}  +
     w_{-1}^{\prime\dagger} 
     \widehat{w}_0^{\vphantom\dagger}                       &(3.3b)\cr
I &= w_{+1}^{\prime\dagger} w_{+1}^{\prime\vphantom\dagger} +
     \widehat{w}_0^{\dagger}
     \widehat{w}_0^{\vphantom{\dagger}} + 
     w_{-1}^{\dagger} w_{-1}^{\vphantom{\dagger}}           &(3.3c)\cr
0 &= w_{+1}^{\dagger}
     \widehat{w}_0^{\vphantom{\dagger}} +
     w_0^{\dagger} w_{-1}^{\vphantom{\dagger}}.             &(3.3d)\cr
}
$$
Constraint $(3.2a)$ is automatically satisfied but, again assuming
that $\cos\rho \not= 0 \not= \cos\rho'$, $(3.3a)$ requires
$\sin(\theta - \theta') = 0$, so we set $\theta' \equiv \theta$.  
Using the form (2.4) for $\widehat{w}_0$, we observe that the 
constraints $(3.2d)$ and $(3.3d)$ are the same as $(2.2b)$ and 
$(2.3b)$, so the $y_i$ must satisfy (2.5) and (2.6).  Constraint
$(3.2b)$ requires that
$$
\eqalign{
y_1 &=   \sin\theta \sin\rho'                                      \cr
y_3 &= -i\cos\theta \sin\rho',                                     \cr
}                                                           \eqno(3.4)
$$
which is consistent with (2.6), just as (2.5) is with $(3.3b)$.  
Combining (2.4), (2.5) and (3.4) we find 
$$
\widehat{w}_0 = \widehat{w}_0(\rho',\theta,\rho) 
 := \pmatrix{  \sin\theta \sin\rho' & -i\cos\theta \sin\rho \cr
             -i\cos\theta \sin\rho' &   \sin\theta \sin\rho \cr
            },                                              \eqno(3.5)
$$
which also satisfies the remaining (normalization) constraints in
(3.2) and (3.3).

The arbitrary phase degree of freedom in the Type~I boundary condition 
is not present in (3.5), but as anticipated, this Type~I inhomogeneity
describes a change in the coupling constant $\rho$, the mass $\theta$ 
being held fixed across the inhomogeneity.  The locus of the 
inhomogeneity is quite precise:  a left moving particle from $x = 0$ 
obeys the `primed' rules, while a right moving particle obeys the
`unprimed' ones.

\medskip
\noindent{\bf 4.  Type II inhomogeneities and boundary conditions}
\nobreak

\nobreak
\noindent The form (3.1) of the Type~I inhomogeneity partitions the 
evolution matrix $U$ into two pieces across an antidiagonal through
the $\widehat{w}_0$ block (inside the block the partition runs between 
the two columns).  We might also consider an inhomogeneity which 
partitions $U$ across an antidiagonal through a pair of $w_{-1}$ and 
$w_{+1}$ blocks.  Such a {\sl Type~II\/} evolution rule inhomogeneity 
has the form:
$$
U :=
\pmatrix{
\ddots
&&&&&&&
\cr
& w_{-1}' 
& w_{ 0}' 
& w_{+1}' 
&&&&
\cr
&
& w_{-1}' 
& w_{ 0}' 
& \widehat{w}_{+1}
&&&
\cr
&&
& \widehat{w}_{-1}^{\vphantom\prime}
& w_{ 0}^{\vphantom\prime} 
& w_{+1}^{\vphantom\prime}
&&
\cr
&&&
& w_{-1}^{\vphantom\prime}
& w_{ 0}^{\vphantom\prime} 
& w_{+1}^{\vphantom\prime}
&
\cr
&&&&&&
& \ddots
\cr
},                                                          \eqno(4.1)
$$
where again $w_i = w_i(\rho,\theta)$ and $w'_i = w_i(\rho',\theta')$ 
are defined by (1.3) but with {\it a priori\/} different parameters.

The unitarity conditions $UU^{\dagger} = I = U^{\dagger}U$ impose even
more constraints in this more complicated situation:
$$
\eqalignno{
I &= w_{-1}^{\prime\vphantom\dagger} w_{-1}^{\prime\dagger} +
     w_0^{\prime\vphantom\dagger} w_0^{\prime\dagger} +
     \widehat{w}_{+1}^{\vphantom\dagger} 
     \widehat{w}_{+1}^{\dagger}                             &(4.2a)\cr
0 &= w_0^{\prime\vphantom\dagger} \widehat{w}_{-1}^{\dagger} +
     \widehat{w}_{+1}^{\vphantom\dagger} w_0^{\dagger}      &(4.2b)\cr
0 &= \widehat{w}_{+1}^{\vphantom\dagger} w_{-1}^{\dagger}   &(4.2c)\cr
0 &= \widehat{w}_{-1}^{\vphantom\dagger} 
     w_{+1}^{\prime\dagger}                                 &(4.2d)\cr
I &= \widehat{w}_{-1}^{\vphantom\prime} \widehat{w}_{-1}^{\dagger} +
     w_0^{\vphantom\dagger} w_0^{\dagger} +
     w_{+1}^{\vphantom\dagger} w_{+1}^{\dagger}             &(4.2e)\cr
\noalign{\hbox{and}}
I &= \widehat{w}_{+1}^{\dagger}
     \widehat{w}_{+1}^{\vphantom\dagger} +
     w_0^{\dagger} w_0^{\vphantom\dagger} +
     w_{-1}^{\dagger} w_{-1}^{\vphantom\dagger}             &(4.3a)\cr
0 &= \widehat{w}_{+1}^{\dagger} w_0^{\prime} +
     w_0^{\dagger} \widehat{w}_{-1}^{\vphantom\dagger}      &(4.3b)\cr
0 &= w_{-1}^{\prime\dagger} 
     \widehat{w}_{+1}^{\vphantom\prime}                     &(4.3c)\cr
0 &= w_{+1}^{\dagger} \widehat{w}_{-1}^{\vphantom\dagger}   &(4.3d)\cr
I &= w_{+1}^{\prime\dagger} w_{+1}^{\prime\vphantom\dagger} +
     w_0^{\prime\dagger} w_0^{\prime\vphantom\dagger} +
     \widehat{w}_{-1}^{\dagger} 
     \widehat{w}_{-1}^{\vphantom\prime}.                    &(4.3e)\cr
}
$$
Suppose the inhomogeneity matrices have the most general forms:
$$
\widehat{w}_{-1} := \pmatrix{ x_1 & x_2 \cr
                              x_3 & x_4 \cr
                            }
\qquad\hbox{and}\qquad
\widehat{w}_{+1} := \pmatrix{ z_1 & z_2 \cr
                              z_3 & z_4 \cr
                            }.                              \eqno(4.4)
$$
Then, assuming $\cos\rho \not= 0$, constraint $(4.2c)$ requires
$z_2 = 0 = z_4$.  Similarly, assuming $\cos\rho' \not= 0$, constraint
$(4.2d)$ requires $x_1 = 0 = x_3$.  Thus the inhomogeneity matrices
have the same advection/scattering interpretation as in the 
homogeneous situation.

Now constraints $(4.3c)$ and $(4.3d)$ imply that
$$
\eqalignno{
z_3 &= iz_1 \tan\theta'                                      &(4.5)\cr
\noalign{\hbox{and}}
x_2 &= ix_4 \tan\theta,                                      &(4.6)\cr
}
$$
respectively.  Imposing the normalization constraint $(4.2a)$ we find 
that 
$$
z_1 = e^{i\zeta} \cos\rho' \cos\theta'.                     \eqno(4.7)
$$
Then imposing the normalization constraint $(4.3a)$ implies 
$\cos^2\rho = \cos^2\rho'$, so we set $\rho' \equiv \rho$.  Combining
(4.4), (4.5) and (4.7) gives
$$
\widehat{w}_{+1}^{\vphantom\prime} = e^{i\zeta} w_{+1}^{\prime}.
                                                            \eqno(4.8)
$$
Similarly, imposing the normalization constraint $(4.2e)$ we find that
$$
x_4 = e^{i\chi} \cos\rho \cos\theta.                        \eqno(4.9)
$$
Combining (4.4), (4.6) and (4.9) gives
$$
\widehat{w}_{-1} = e^{i\chi} w_{-1},                       \eqno(4.10)
$$
which also satisfies the last normalization constraint $(4.3e)$.  The
two remaining constraints $(4.2b)$ and $(4.3b)$ require only that
$\chi \equiv -\zeta$ (mod $2\pi$), which can thence be set to 0 by a 
unitary transformation.  Thus (4.8) and (4.10) become
$$
\widehat{w}_{-1} = w_{-1}(\rho,\theta) 
\qquad\hbox{and}\qquad
\widehat{w}_{+1} = w_{+1}(\rho,\theta').                   \eqno(4.11)
$$

Just as the Type~I inhomogeneity described by (3.1) and (3.5) 
specializes to a Type~I boundary condition described by (2.1) and 
(2.8) when $\cos\rho' = 0$ so that there is no advection to the left 
of $x = 0$, the Type~II inhomogeneity described by (4.1) and (4.11) 
specializes to a boundary condition when $\cos\theta' = 0$.  In this 
situation, when a left moving particle at $x = 1$ advects to $x = 0$ 
it scatters to the right, while a right moving particle at $x = 0$ 
which remains at $x = 0$ also scatters to the right---a particle 
initially at $x > 0$ or at $x = 0$ and right moving has no amplitude 
to be at $x < 0$ or at $x = 0$ and left moving at any subsequent 
timestep.

This is a special case of the {\sl Type~II\/} boundary condition 
which we expect to be characterized by nontrivial phases, just as is
the Type~I boundary condition.  The `primed' parameters satisfy
$\rho' \equiv \rho$ and $\cos\theta' = 0$, so generalizing the Type~II
inhomogeneity by multiplicative phases suggests
$$
U :=
\pmatrix{
 e^{i\upsilon}w_{0}^{\prime} 
 & e^{i\zeta}w_{+1}^{\prime} &        &        \cr
           \overline{w}_{-1} & w_{ 0} & w_{+1} &      \cr
                             & w_{-1} & w_{ 0} &      \cr
                             &        &        &\ddots\cr
},                                                         \eqno(4.12)
$$
where
$$
\overline{w}_{-1} 
 = \cos\rho \pmatrix{ 0 & i e^{i\chi_1} \sin\theta \cr
                      0 &   e^{i\chi_2} \cos\theta \cr
                    }                                      \eqno(4.13)
$$
is a generalization of (4.10).  Then the unitarity conditions impose 
constraints on the phase angles $\chi_1$, $\chi_2$, $\upsilon$ and 
$\zeta$ {\it via\/} (4.2) and (4.3).  Constraint $(4.2c)$ is 
automatically satisfied, while $(4.3d)$ requires 
$\chi_1 \equiv \chi_2 =: \chi$ (mod $2\pi$).  This means that the 
normalization constraints are necessarily satisfied so the only 
remaining constraints are $(4.2b)$ and $(4.3b)$.  These are satisfied 
provided $\upsilon \equiv \chi + \zeta$ (mod $2\pi$).  Finally, up to 
unitary equivalence we may set $\chi = 0$, so the most general Type~II 
boundary condition is defined by:
$$
U =
\pmatrix{
   e^{i\zeta}w_{0}^{\prime} 
 & e^{i\zeta}w_{+1}^{\prime} &        &        \cr
             w_{-1} & w_{ 0} & w_{+1} &        \cr
                    & w_{-1} & w_{ 0} &        \cr
                    &        &        & \ddots \cr
},                                                         \eqno(4.14)
$$
where $w_i^{\prime} = w_i(\rho,0)$.  Just as does the Type~I boundary 
condition, the Type~II boundary condition has one phase degree of 
freedom.

\medskip
\noindent{\bf 5.  Type III boundary conditions}
\nobreak

\nobreak
\noindent The two types of inhomogeneities we have found reflect the 
$\rho \longleftrightarrow \theta$ duality evident in the dispersion 
relation (6.2) discussed in [\qmI]:  The Type~I inhomogeneity has 
constant $\theta$ and discontinuity in $\rho$ while the Type~II 
inhomogeneity has constant $\rho$ and discontinuity in $\theta$.  
Suppose we wish to change both $\rho$ and $\theta$.  This is clearly 
possible using a Type~I inhomogeneity to change $\rho$ followed by a 
Type~II inhomogeneity to change $\theta$, provided the discontinuities 
are sufficiently far apart that the constraints (3.2), (3.2), (4.2) 
and (4.3) do not overlap.  In fact, the discontinuities can be 
adjacent:  it is straightforward to verify that the evolution matrix
$$
U :=
\pmatrix{
\ddots
&&&&&&&
\cr
& w_{-1}' 
& w_{ 0}' 
& w_{+1}' 
&&&&
\cr
&
& w_{-1}' 
& \widehat{w}_{ 0} 
& \widehat{w}_{+1}
&&&
\cr
&&
& \widehat{w}_{-1}^{\vphantom\prime}
& w_{ 0}^{\vphantom\prime} 
& w_{+1}^{\vphantom\prime}
&&
\cr
&&&
& w_{-1}^{\vphantom\prime}
& w_{ 0}^{\vphantom\prime} 
& w_{+1}^{\vphantom\prime}
&
\cr
&&&&&&
& \ddots
\cr
}                                                           \eqno(5.1)
$$
is unitary for $\widehat{w}_{-1} = w_{-1}(\rho,\theta)$ and 
$\widehat{w}_{+1} = w_{+1}(\rho,\theta')$ as in (4.11) and 
$\widehat{w}_0 = \widehat{w}_0(\rho',\theta',\rho)$.  The evolution 
matrix (5.1) describes a system in which the parameters $\rho'$ and 
$\theta'$ change to $\rho$ and $\theta$ across the inhomogeneity.

While (5.1) does not describe a new type of inhomogeneity as it 
is composed of a Type~I and Type~II pair, our experience with boundary
conditions in the previous sections suggest that there may be an 
analogous {\sl Type~III\/} boundary condition which has extra phase 
degrees of freedom.  Suppose
$$
U :=
\pmatrix{
 \overline{w}_{0} & \overline{w}_{+1} &        &      \cr
           \overline{w}_{-1} & w_{ 0} & w_{+1} &      \cr
                             & w_{-1} & w_{ 0} &      \cr
                             &        &        &\ddots\cr
},                                                          \eqno(5.2)
$$
where the $w_i$ are given by (1.3), $\overline{w}_{-1}$ is given by
(4.13), $\overline{w}_{+1}$ is generalized from (4.11):
$$
\overline{w}_{+1} 
 = \cos\rho \pmatrix{  e^{i\zeta_1} \cos\theta' & 0 \cr
                     i e^{i\zeta_2} \sin\theta' & 0 \cr
                    },
$$
and $\overline{w}_0$ is the same as in (2.8) with $\theta$ replaced 
by $\theta'$ and also with additional phase factors:
$$
\overline{w}_0 
 = \pmatrix{  e^{i\upsilon_1} \sin\theta' & 
            -ie^{i\upsilon_2} \cos\theta \sin\rho \cr
            -ie^{i\upsilon_3} \cos\theta' &   
              e^{i\upsilon_4} \sin\theta \sin\rho \cr
           }.
$$

In this case, the unitarity conditions require
$$
\eqalignno{
I &= \overline{w}_0^{\vphantom\dagger} \overline{w}_0^{\dagger} +
     \overline{w}_{+1}^{\vphantom\dagger} \overline{w}_{+1}^{\dagger}
                                                            &(5.3a)\cr
0 &= \overline{w}_0^{\vphantom\dagger} \overline{w}_{-1}^{\dagger} +
     \overline{w}_{+1}^{\vphantom\dagger} w_0^{\dagger}     &(5.3b)\cr
0 &= \overline{w}_{+1}^{\vphantom\dagger} w_{-1}^{\dagger}  &(5.3c)\cr
I &= \overline{w}_{-1}^{\vphantom\dagger} \overline{w}_{-1}^{\dagger}
     + w_0^{\vphantom\dagger} w_0^{\dagger} 
     + w_{+1}^{\vphantom\dagger} w_{+1}^{\dagger}           &(5.3d)\cr
\noalign{\hbox{and}}
I &= \overline{w}_0^{\dagger} \overline{w}_0^{\vphantom\dagger} +
     \overline{w}_{-1}^{\dagger} \overline{w}_{-1}^{\vphantom\dagger} 
                                                            &(5.4a)\cr
0 &= \overline{w}_{+1}^{\dagger} \overline{w}_0^{\vphantom\dagger} +
     w_0^{\dagger} \overline{w}_{-1}^{\vphantom\dagger}     &(5.4b)\cr
0 &= w_{+1}^{\dagger} \overline{w}_{-1}^{\vphantom\dagger}  &(5.4c)\cr
I &= \overline{w}_{+1}^{\dagger} \overline{w}_{+1}^{\phantom\dagger}
     + w_0^{\dagger} w_0^{\vphantom\dagger}
     + w_{-1}^{\dagger} w_{-1}^{\vphantom\dagger}.          &(5.4d)\cr
}
$$
Constraints $(5.3c)$ and $(5.4c)$ are the same as $(4.2c)$ and 
$(4.3b)$, respectively, so they have the same consequences as in the 
case of the Type~II boundary condition:  $(5.3c)$ is satisfied
automatically, while $(5.4c)$ requires 
$\chi_1 \equiv \chi_2 =: \chi$ (mod $2\pi$).  Next, $(5.3b)$ implies 
$\upsilon_2 \equiv \chi + \zeta_1$ (mod $2\pi$) and
$\upsilon_4 \equiv \chi + \zeta_2$ (mod $2\pi$).  Constraint $(5.4b)$ 
requires $\upsilon_3 - \upsilon_1 \equiv \zeta_2 - \zeta_1 =: \delta$
(mod $2\pi$), whereupon the remaining (normalization) constraints in 
(5.3) and (5.4) are automatically satisfied.  Combining these results 
and setting $\upsilon := \upsilon_1$, $\zeta := \zeta_1$ gives
$$
\setbox1=\hbox{%
$
\overline{w}_{-1} 
 = e^{i\chi}  \cos\rho\pmatrix{0 & i\sin\theta  \cr
                               0 &  \cos\theta  \cr
                              }
\qquad
\overline{w}_{+1} 
 = e^{i\zeta} \cos\rho\pmatrix{             \cos\theta' & 0 \cr
                               ie^{i\delta} \sin\theta' & 0 \cr
                              }
$
}
\eqalign{
\overline{w}_{-1} 
 = e^{i\chi}  \cos\rho\pmatrix{0 & i\sin\theta  \cr
                               0 &  \cos\theta  \cr
                              }
\qquad
\overline{w}_{+1} 
 = e^{i\zeta} \cos\rho\pmatrix{             \cos\theta' & 0 \cr
                               ie^{i\delta} \sin\theta' & 0 \cr
                              }                                    \cr
\hbox to\wd1{\hfil%
$
\overline{w}_0  
 = \pmatrix{  e^{i\upsilon} \sin\theta' & 
            -ie^{i(\chi + \zeta)} \cos\theta'\sin\rho \cr
            -ie^{i(\upsilon + \delta)} \cos\theta' &   
              e^{i(\chi + \zeta + \delta)} \sin\theta'\sin\rho \cr
           }
$,
\hfil}                                                             \cr
} 
$$
for the weights in (5.2).  We may set $\chi = 0 = \delta$ by a unitary
transformation, so the most general Type~III boundary condition, up to 
unitary equivalence, is given by 
$\overline{w}_{-1} = w_{-1}(\rho,\theta)$, 
$\overline{w}_{+1} = e^{i\zeta} w_{+1}(\rho,\theta')$, and
$$
\overline{w}_0 
 = \pmatrix{  e^{i\upsilon} \sin\theta' & 
            -ie^{i\zeta} \cos\theta'\sin\rho \cr
            -ie^{i\upsilon} \cos\theta' &   
              e^{i\zeta} \sin\theta'\sin\rho \cr
           }.
$$
As expected, in addition to $\theta'$ there are two phase angle 
degrees of freedom:  $\upsilon$ and $\zeta$.

\medskip
\noindent{\bf 6.  Plane waves near a boundary}
\nobreak

\nobreak
\noindent The global evolution matrices (2.1), (4.14) and (5.2) 
describe unitary evolution of a single particle in the presence of a 
boundary of Type~I, II, or III, respectively.  Away from the boundary
the local evolution is still given by (1.1) and (1.3), so the one
particle plane waves 
$$
\psi^{(k,\epsilon)}(x) = \psi^{(k,\epsilon)}(0) e^{ikx}     \eqno(6.1)
$$
we found in [\qmI] still evolve, locally, by multiplication by 
$e^{-i\epsilon\omega}$ at each time step, where $\omega$ satisfies the 
dispersion relation
$$
\cos\omega = \cos k \cos\theta \cos\rho + \sin\theta \sin\rho
                                                            \eqno(6.2)
$$
and $\epsilon \in \{\pm1\}$.  In fact, any linear combination
$$
\psi^{(k,\epsilon)}(x) + A \psi^{(-k,\epsilon)}(x)          \eqno(6.3)
$$
evolves locally by phase multiplication as both $k$ and $-k$ satisfy
(6.2) with the same frequency $\omega$.

Consider the Type~I boundary condition at $x = 0$ and suppose there is 
an eigenfunction $\psi^{(\omega)}(x)$ of the form (6.3), which should 
be interpreted as a linear combination of incident and reflected plane 
waves with relative amplitude $A$, just as in the situation of 
scattering off a potential step considered in [\lgbu].  Then
$$
\overline{w}_0 \psi^{(\omega)}(0) + w_{+1} \psi^{(\omega)}(1)
 = e^{-i\omega} \psi^{(\omega)}(0).                         \eqno(6.4)
$$
The linear combination (6.3) is well defined for $x < 0$ and
$$
w_{-1} \psi^{(\omega)}(-1) + 
   w_0 \psi^{(\omega)}(0) + 
w_{+1} \psi^{(\omega)}(+1) 
 = e^{-i\omega} \psi^{(\omega)}(0)                          \eqno(6.5)
$$
for any $A \in \C$, so subtracting (6.5) from (6.4) gives
$$
(\overline{w}_0 - w_0) \psi^{(\omega)}(0) 
 = w_{-1} \psi^{(\omega)}(-1).                              \eqno(6.6)
$$
Using (1.3), (2.8) and (6.3) in (6.6) we find
$$
A = -{(e^{i\upsilon} - \sin\rho)   \psi^{(k,\epsilon)}_{-1}(0)
            - i e^{-ik} \cos\rho \,\psi^{(k,\epsilon)}_{+1}(0)
      \over
      (e^{i\upsilon} - \sin\rho)   \psi^{(-k,\epsilon)}_{-1}(0)
             - i e^{ik} \cos\rho \,\psi^{(-k,\epsilon)}_{+1}(0)
     }                                                      \eqno(6.7)
$$
where 
$$
\psi^{(k,\epsilon)}(0) 
:= \pmatrix{i \sin\rho \cos\theta - i e^{-ik} \cos\rho \sin\theta  \cr
              \sin\rho \sin\theta +   e^{ik}  \cos\rho \cos\theta 
                                  -   e^{-i\epsilon\omega}         \cr
           }                                                \eqno(6.8)
$$
is the (unnormalized) eigenvector of $D(k)$ in [\qmI].  That is, with 
$A$ given by (6.7), the linear combination (6.3) is an eigenfunction
satisfying the Type~I boundary condition.

The more complicated Type~II and III boundary conditions require 
modifications to the linear combination of plane waves (6.3) near the
boundary.  Consider the Type~II boundary condition and suppose
$$
\eqalignno{
\psi^{(\omega)}(x) 
 &:= \psi^{(k,\epsilon)}(x) + A \psi^{(-k,\epsilon)}(x)     
\qquad\hbox{for}\qquad
x \ge 1                                                      &(6.9)\cr
\noalign{\hbox{and}}
\psi^{(\omega)}_{-1}(0) &:= 0,                              &(6.10)\cr
}
$$
where the latter condition follows from the discussion preceding 
(4.12).  At $x = 1$ the same argument as in the Type~I boundary case
gives
$$
\overline{w}_{-1} \psi^{(\omega)}(0) 
 = w_{-1} \bigl(\psi^{(k,\epsilon)}(0) + A \psi^{(-k,\epsilon)}(0) 
          \bigr)                                           \eqno(6.11)
$$
which implies
$$
\psi^{(\omega)}_{+1}(0) 
 := \psi^{(k,\epsilon)}_{+1}(0) + A \psi^{(-k,\epsilon)}_{+1}(0). 
                                                           \eqno(6.12)
$$
Applying (4.14) to the eigenfunction $\psi^{(\omega)}(x)$ at $x = 0$
gives
$$
e^{i\zeta} w'_0 \psi^{(\omega)}(0) + 
e^{i\zeta} w'_{+1} \psi^{(\omega)}(1) 
 = e^{-i\omega} \psi^{(\omega)}(0).                        \eqno(6.13)
$$
Using the expressions for $w'_i$ with $\rho' \equiv \rho$, 
$\cos\theta' = 0$ and (6.9), (6.10) and (6.12) in (6.13) we find
$$
A = -{\bigl(e^{i\zeta} \sin\rho - e^{-i\omega}
      \bigr) \psi^{(k,\epsilon)}_{+1}(0)
       + i e^{i(\zeta + k)} \cos\rho \,\psi^{(k,\epsilon)}_{-1}(0)
      \over
      \bigl(e^{i\zeta} \sin\rho - e^{-i\omega}
      \bigr) \psi^{(-k,\epsilon)}_{+1}(0)
       + i e^{i(\zeta - k)} \cos\rho \,\psi^{(-k,\epsilon)}_{-1}(0)
     }.                                                    \eqno(6.14)
$$
Thus (6.9) with $A$ given by (6.14), (6.10) and (6.12) define an
eigenfunction satisfying the Type~II boundary condition.

To find the eigenfunctions for the Type~III boundary condition, we 
still suppose they satisfy (6.9), but not (6.10).  Since 
$\overline{w}_{-1}$ is the same as for the Type~II boundary, (6.11)
still implies (6.12).  Now applying (5.2) to the eigenfunction
$\psi^{(\omega)}(x)$ at $x = 0$ gives
$$
\overline{w}_0 \psi^{(\omega)}(0) +
\overline{w}_{+1} \psi^{(\omega)}(1)
 = e^{-i\omega} \psi^{(\omega)}(0),
$$
which comprises a pair of linear equations for 
$\psi^{(\omega)}_{-1}(0)$ and $A$.  These equations can be solved to
give the eigenfunctions for the Type~III boundary conditions, although
we will not need the explicit solution here.

\medskip
\noindent{\bf 7.  Plane waves on finite lattices}
\nobreak

\nobreak
\noindent With only one boundary, \eg, $L = \N$ as we were 
considering implicitly in the previous section, the wave number can 
take any value in the interval $-\pi < k \le \pi$ and the
frequency/energy spectrum is continuous with range 
$\theta - \rho \le |\omega| \le \pi - (\theta + \rho)$ (assuming
$0 \le \rho \le \theta \le \pi/2$) determined by the dispersion 
relation (6.2).  On finite lattices, however, the spectra are discrete 
and are determined by the two boundary conditions.  Consider the case 
of two Type~I boundary conditions on a lattice of cardinality $N$.  
The weights in the boundary condition at $x = N-1$ are the parity 
transforms of those in (6.6):
$$
P(\overline{w}_0 - w_0)P^{-1} \psi^{(\omega)}(N-1)
 = Pw_{-1}P^{-1} \psi^{(\omega)}(N),
$$
where
$$
P := \pmatrix{0 & 1 \cr
              1 & 0 \cr
             }.
$$
This gives a second constraint on $A$:
$$
A = -e^{-2ik(N-1)}
    {(e^{i\upsilon} - \sin\rho)   \psi^{(k,\epsilon)}_{+1}(0)
            - i e^{ik} \cos\rho \,\psi^{(k,\epsilon)}_{-1}(0)
     \over
     (e^{i\upsilon} - \sin\rho)   \psi^{(-k,\epsilon)}_{+1}(0)
           - i e^{-ik} \cos\rho \,\psi^{(-k,\epsilon)}_{-1}(0)
     }                                                      \eqno(7.1)
$$
which must be consistent with (6.7).  To see how this determines the
discrete spectrum, let $\upsilon = 0 = \rho$.  Then (6.7) becomes
$$
A = -e^{-2ik} { e^{ik} \cos\theta - e^{-i\epsilon\omega} + \sin\theta
               \over
               e^{-ik} \cos\theta - e^{-i\epsilon\omega} + \sin\theta
              }                                             \eqno(7.2)
$$
and (7.1) becomes
$$
A = -e^{-2ik(N-1)} 
              { e^{ik} \cos\theta - e^{-i\epsilon\omega} - \sin\theta
               \over
               e^{-ik} \cos\theta - e^{-i\epsilon\omega} - \sin\theta
              }.                                            \eqno(7.3)
$$
Setting the right hand sides of (7.2) and (7.3) to be equal and using 
the dispersion relation (6.2) to eliminate $\omega$, we find, after 
some algebra,
$$
e^{-2i(N-2)k} (\sin\theta - i\sin k \cos\theta)
 = \sin\theta + i\sin k \cos\theta.                         \eqno(7.4)
$$
Supposing $k$ to be real, the right hand side of (7.4) is the complex
conjugate of the parenthesized expression on the left hand side, which
implies that
$$
\tan\bigl((N-2)k\bigr) + \sin k \cot\theta = 0.             \eqno(7.5)
$$
The left hand side of (7.5) has poles at $k = (n + 1/2) \pi/(N-2)$,
$n \in \Z$, between each pair of which there must be a root of the
equation.  Thus (7.5) has $N-1$ roots in the interval 
$0 \le k \le \pi$, giving $N-1$ distinct values for eigenfrequencies 
in the range $\theta - \rho \le \omega \le \pi - \theta - \rho$ 
(assuming $0 \le \rho \le \theta \le \pi/2$).  But $U$ is a 
$2N \times 2N$ matrix so it must have $2N$ eigenvalues $e^{-i\omega}$.  
Figures~1 and 2 show the results of computing the eigenvalues of $U$ 
numerically for $N = 16$:  the eigenfrequencies are plotted as 
functions of the Type~I boundary parameter $\upsilon$, set to the same 
value at each boundary.  Notice that while most of the 
eigenfrequencies lie in the expected intervals, there are four which, 
over parts of the range of $\upsilon$, do not.

\midinsert
$$
\epsfxsize=\halfwidth\epsfbox{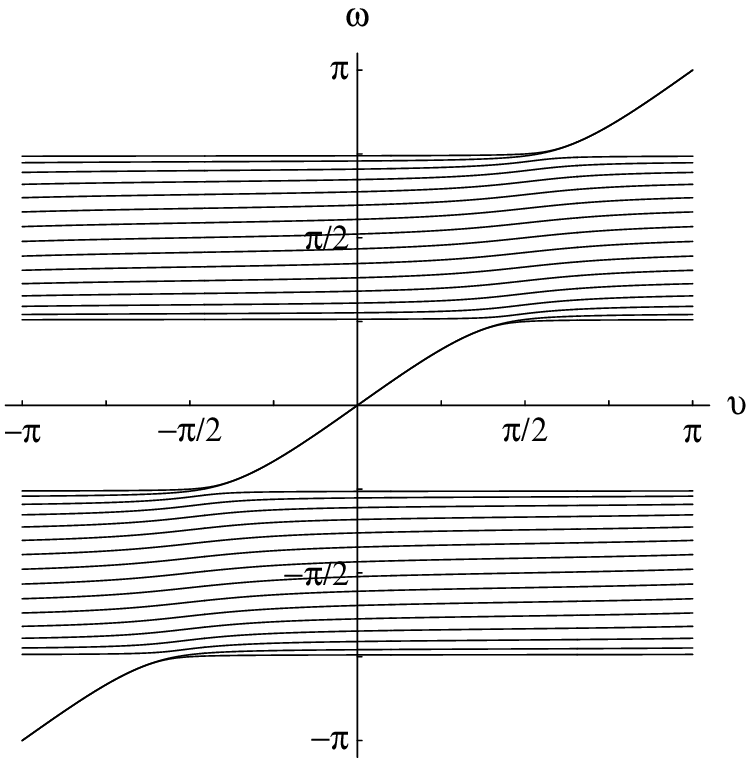}\hskip\chasm%
\epsfxsize=\halfwidth\epsfbox{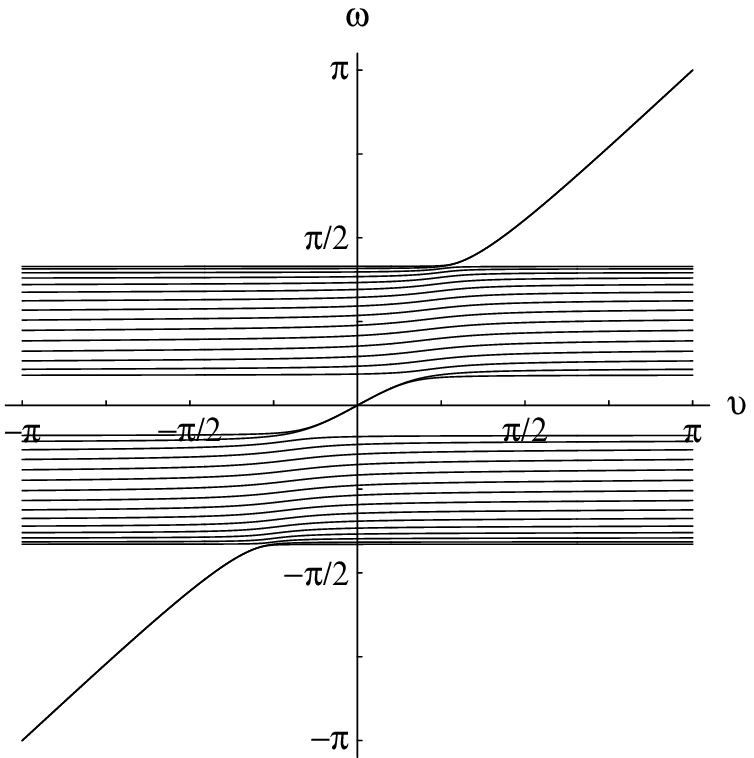}
$$
\hbox to\hsize{%
\vbox{\hsize=\halfwidth\eightpoint{%
\noindent{\bf Figure~1}.  The eigenfrequencies $\omega$ of $U$ for a 
lattice of size $N = 16$ with $\rho = 0$, $\theta = \pi/4$ and two
Type~I boundary conditions with the same parameter $\upsilon$.
}}
\hfill%
\vbox{\hsize=\halfwidth\eightpoint{%
\noindent{\bf Figure~2}.  The same situation as in Figure~1 but with
parameters $\rho = \pi/4$, $\theta = \pi/3$.  In both cases there are 
two eigenvalues with $\omega \approx 0$ when $\upsilon = 0$ and two 
with $\omega \approx \pi$ when $\upsilon = \pi$.
}}}
\endinsert

To understand the origin of these unexpected eigenfrequencies, let us 
reconsider (7.4) and suppose that $k$ has a nonzero imaginary part.
Then for large $N$ and the correct sign of $k$, the left hand side of
(7.4) becomes arbitrarily small.  So, if there were such a $k$ which
caused the right hand side of (7.4) to vanish, it would provide an
additional root.  Solving
$$
0 = \sin\theta + {1\over 2}(e^{ik} - e^{-ik}) \cos\theta,
$$
we find 
$$
e^{ik} = -\tan\theta \pm \sec\theta.                        \eqno(7.6)
$$
The negative root in (7.6) makes the norm $|e^{-ik}| \le 1$ for 
$0 \le \theta \le \pi/2$; furthermore, it satisfies the dispersion 
relation (6.2) with $\omega = 0$.  Thus in the $N \to \infty$ limit,
1 is an eigenvalue of $U$ with multiplicity two.  For finite $N$ these 
extra eigenfrequencies split, finely, and are only very close to 0.  
As we see in Figures~1 and 2, as $\upsilon$ changes away from 0, the
splitting increases and the eigenfrequencies move into the range
associated with real wave numbers.  An analogous discussion explains
the pair of eigenfrequencies near $\pi$ at $\upsilon = \pi$.  The
eigenfunctions having these eigenfrequencies corresponding to wave
numbers with nonzero imaginary part are, of course, not plane waves;
rather, each describes the state of a `low' energy particle which is
`trapped' at the boundaries, with exponentially decreasing amplitude
to be in the interior of the lattice.

For the case of two Type~II boundary conditions note that {\it per\/} 
the discussion following (4.12) the eigenfunctions of interest are 
those which have vanishing left (right) moving amplitude at the left 
(right) boundary.  Thus when $|L| = N$, there are $2N - 2$ relevant
eigenfunctions and eigenfrequencies.  Figures~3 and 4 show the results
of computing the eigenvalues of $U$ numerically for $N = 16$:  the
eigenfrequencies are plotted as functions of the Type~II boundary
parameter $\zeta$, set to the same value at each boundary.  As in the
Type~I boundary situation, most of the eigenfrequencies lie in the 
ranges corresponding to real wave numbers, although near $\zeta = \pi$
there are four which do not, and which are explained by an analysis 
similar to that of the preceding paragraph.

\topinsert
$$
\epsfxsize=\halfwidth\epsfbox{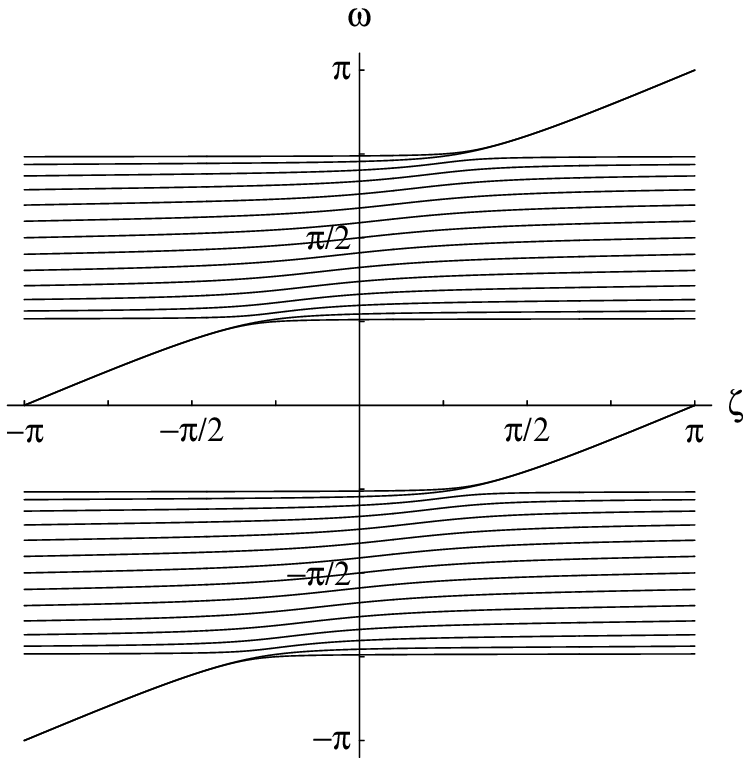}\hskip\chasm%
\epsfxsize=\halfwidth\epsfbox{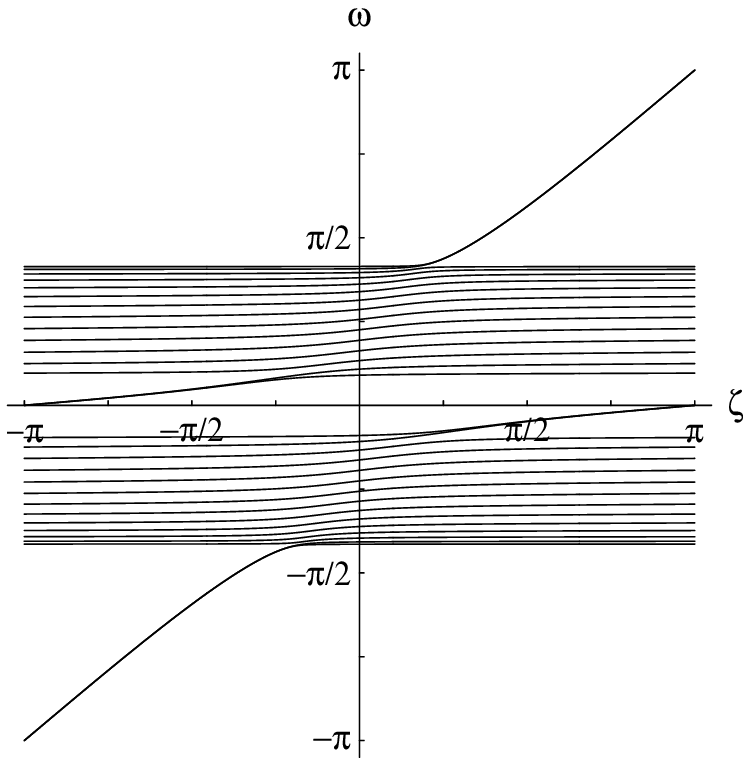}
$$
\hbox to\hsize{%
\vbox{\hsize=\halfwidth\eightpoint{%
\noindent{\bf Figure~3}.  The eigenfrequencies $\omega$ of $U$ for a 
lattice of size $N = 16$ with $\rho = 0$, $\theta = \pi/4$ and two
Type~II boundary conditions with the same parameter $\zeta$.
}}
\hfill%
\vbox{\hsize=\halfwidth\eightpoint{%
\noindent{\bf Figure~4}.  The same situation as in Figure~3 but with
parameters $\rho = \pi/4$, $\theta = \pi/3$.  In both cases there are 
two eigenvalues with $\omega \approx 0$ and two with 
$\omega \approx \pi$ when $\zeta = \pi$.
}}}
\endinsert

Finally, consider the case of two Type~III boundary conditions, again
with equal parameter values.  In this case there is a non-phase 
parameter which can be adjusted, namely $\theta'$.  Figure~5 shows the
eigenfrequencies of $U$ as a function of $\theta'$ for the rule 
defined by $\rho = 0$, $\theta = \pi/4$, with boundary parameters
$\upsilon = 0 = \zeta$.  To separate the eigenvalues we have computed 
them for a lattice of size only $N = 4$.  Figure~6 is similar, but 
the rule parameters are now $\rho = \pi/4$, $\theta = \pi/3$.  In this 
case $N = 8$ and the two eigenfrequencies near 0 are only finely split 
over the whole parameter range.  Notice that in each case there are 
actually six eigenvalues corresponding to imaginary wave numbers.  
Examination of the eigenfunctions shows that the two with 
eigenfrequencies near 0 have amplitudes concentrated in the states 
$|1,-1\rangle$ and $|N-2,+1\rangle$, while the four with 
eigenfrequencies closer to $\pm\pi$ have amplitudes concentrated at 
$x = 0$ and $x = N-1$.

\topinsert
$$
\epsfxsize=\halfwidth\epsfbox{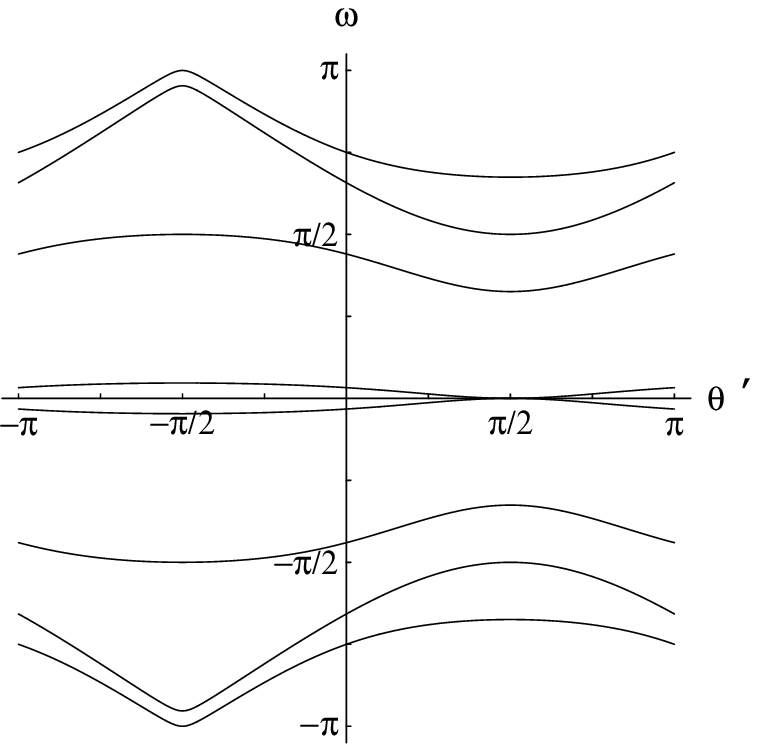}\hskip\chasm%
\epsfxsize=\halfwidth\epsfbox{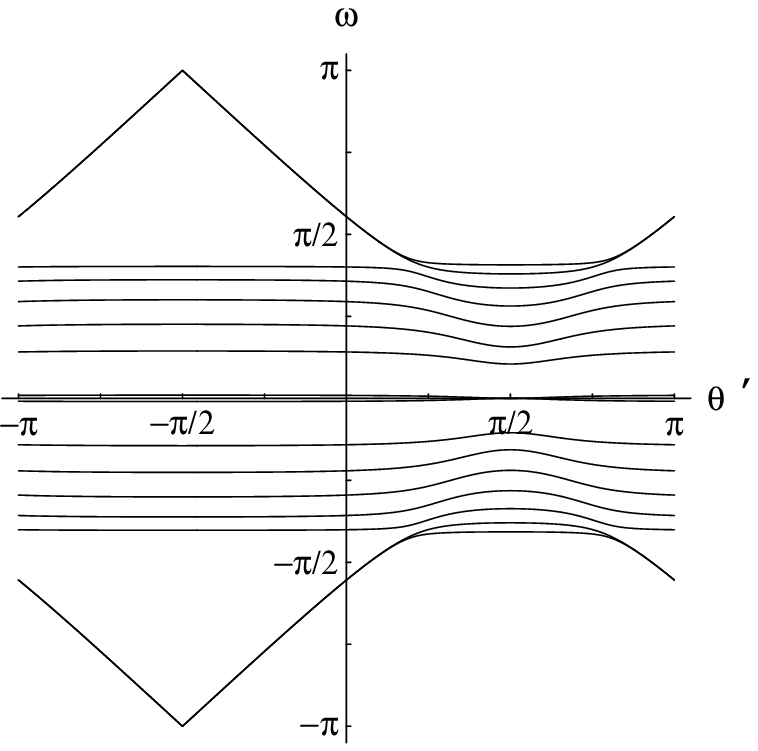}
$$
\hbox to\hsize{%
\vbox{\hsize=\halfwidth\eightpoint{%
\noindent{\bf Figure~5}.  The eigenfrequencies $\omega$ of $U$ for a 
lattice of size $N = 4$ with $\rho = 0$, $\theta = \pi/4$ and two
Type~III boundary conditions with the same parameter $\theta'$ and
both boundary phase angles 0.
}}
\hfill%
\vbox{\hsize=\halfwidth\eightpoint{%
\noindent{\bf Figure~6}.  The same situation as in Figure~5 but with
parameters $\rho = \pi/4$, $\theta = \pi/3$, and $N = 8$.  In both 
cases there are two eigenvalues with $\omega \approx 0$ and four with 
$|\omega| \approx \pi$.
}}}
\endinsert

\medskip
\noindent{\bf 8.  Reflection and refraction of wave packets}
\nobreak

\nobreak
\noindent The physical meaning of the rule inhomogeneities we are 
considering is perhaps most clear in wave packet simulations.  In [\qmI]
we defined binomial wavepackets with localized initial position and
particularized initial wavenumber.  In each of the simulations of this
section the initial wavepacket is built from a plane wave (6.1) and 
(6.8) with $k_0 = \pi/4$, is centered at $x = 16$ and has width 32, on 
the lattice $0 \le x \le 63$.  The peak frequency $\omega_0$ and the 
group velocity depend on the rule parameters $\rho$ and $\theta$ 
through the dispersion relation (6.2).

Let us first consider the reflection of such a wave packet from the 
possible boundaries.  Figure~7 shows the evolution of the wave packet 
with parameters $\rho = 0$ and $\theta = \pi/4$ in the presence of 
Type~I boundary conditions with $\upsilon = 0$.  Reflection from 
Type~II and Type~III boundaries is extremely similar:  in each case 
the significant dispersion of the wave packet at the time of 
interaction with the wall results in a sequence of reflected (smaller) 
wave packets.

\pageinsert
\null\vskip-3\baselineskip
$$
\epsfxsize=\usewidth\epsfbox{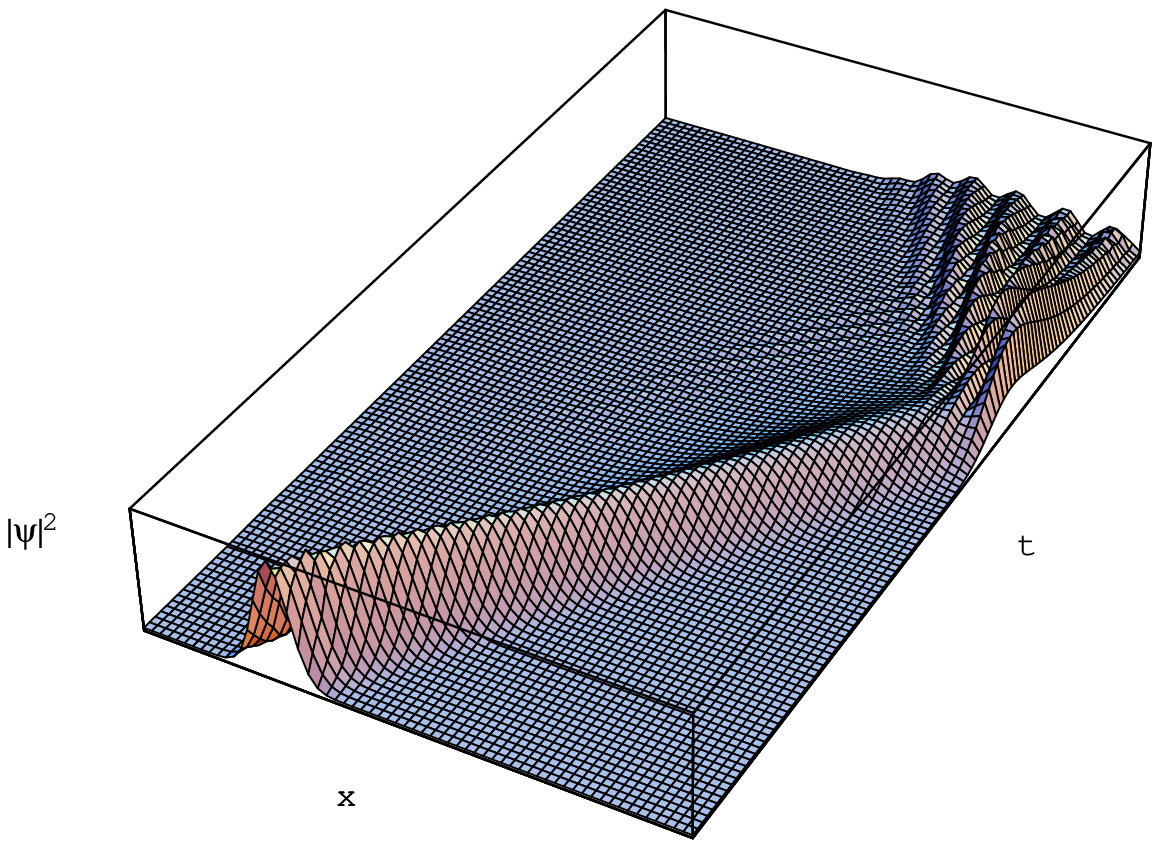}
$$
\vskip-1.5\baselineskip
\eightpoint{%
{\narrower\noindent{\bf Figure~7}.  Evolution of the $k_0 = \pi/4$
wave packet with width 32 for rule parameters $\rho = 0$, 
$\theta = \pi/4$.  The boundaries are both of Type~I with 
$\upsilon = 0$.\par} 
}

\vfill
\null\vskip-3\baselineskip
\vfill
\centerline{({\tenit figure available from author\/})}
\vfill
\vskip-1.5\baselineskip
\eightpoint{%
{\narrower\noindent{\bf Figure~8}.  Evolution of the same wave packet 
as in Figure~7 for rule parameters $\rho = \pi/4 = \theta$.  The 
boundaries are both of Type~II with $\zeta = 0$.\par} 
}
\endinsert

As we learned in [\qmI], a `massless' wave packet disperses more slowly
than a massive one.  In Figure~8 we show the results of a simulation 
of this case:  $\rho = \pi/4 = \theta$ and the boundaries are both of 
Type~II with $\zeta = 0$.  Reflection from Type~I and Type~III 
boundaries is again similar:  in each case the wave packet reflects 
cleanly and suffers little more dispersion than if the boundary had 
not been there.

Now let us consider the effect of rule inhomogenities on wave packet
evolution.  Figure~9 shows the results of a simulation in which there
is a Type~I inhomogeneity at $x = 31$:  the rule parameter 
$\theta$ is constant at $\pi/4$ while $\rho$ is 0 to the left, and 
$\pi/4$ to the right, of the inhomogeneity.  There is both reflection 
and transmission of the wave packet at the inhomogeneity:  the 
reflected wave disperses rapidly which causes an interaction with the
left boundary similar to that shown in Figure~7 while the transmitted
wave packet has little dispersion and evolves much as the wave packet
in Figure~8.

\pageinsert
\null\vskip-3\baselineskip
$$
\epsfxsize=\usewidth\epsfbox{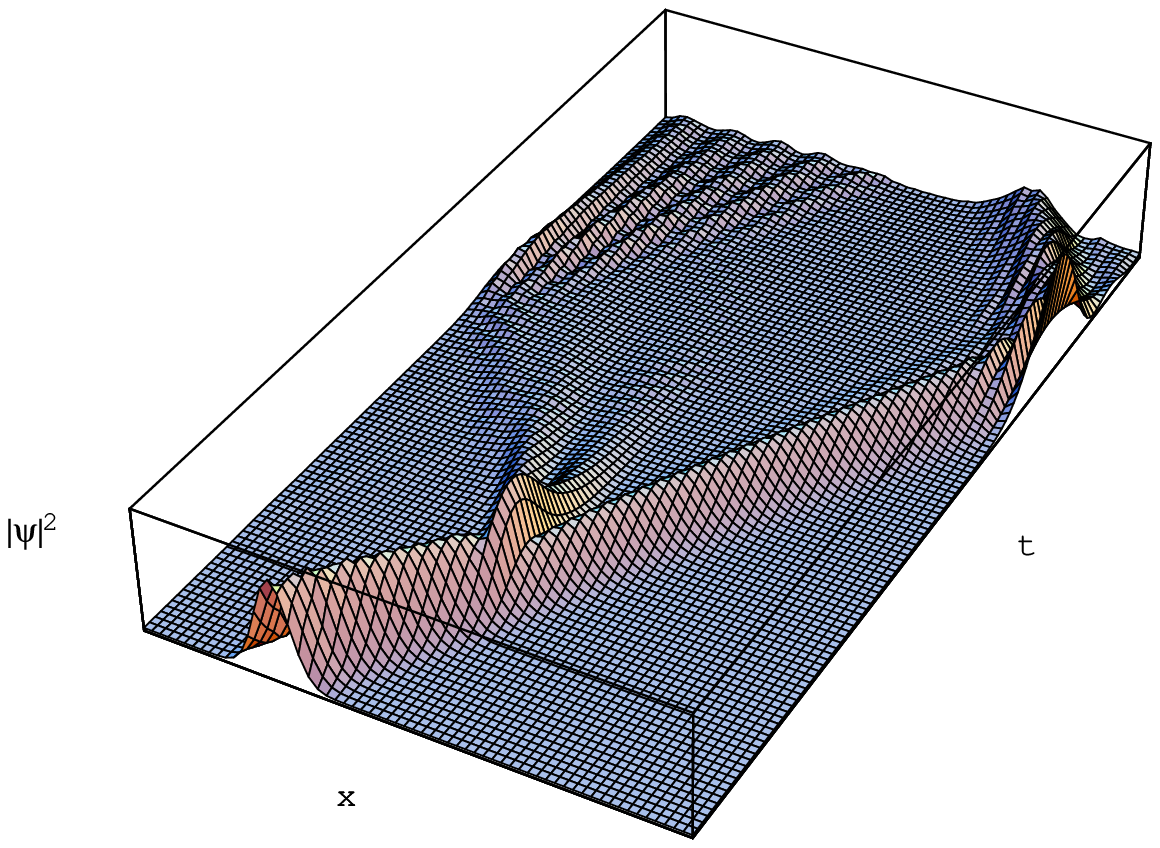}
$$
\vskip-1.5\baselineskip
\eightpoint{%
{\narrower\noindent{\bf Figure~9}.  Evolution of the same wave packet 
as in the previous figures with rule parameters $\theta = \pi/4$ 
everywhere and $\rho = 0$ to the left and $\rho = \pi/4$ to the right
of a Type~I inhomogeneity at $x = 31$.  Both boundaries are of Type~I
with $\upsilon = 0$.\par} 
}

\vfill
\null\vskip-3\baselineskip
\vfill
\centerline{({\tenit figure available from author\/})}
\vfill
\vskip-1.5\baselineskip
\eightpoint{%
{\narrower\noindent{\bf Figure~10}.  Evolution of the same wave packet 
as in the previous figures with rule parameters $\rho = \pi/4$ 
everywhere and $\theta = \pi/4$ to the left and $\theta = \pi/3$ to
the right of a Type~II inhomogeneity at $x = 32$.  Both boundaries are 
of Type~II with $\zeta = 0$.\par} 
}
\endinsert

Next, let us consider the effect of a Type~II inhomogeneity which 
changes the value of $\theta$ from $\pi/4$ to the left of $x = 32$ to
$\pi/3$ to the right.  Figure~10 show the results of a simulation with
$\rho = \pi/4$ everywhere and Type~II boundary conditions.  To the 
left of the inhomogeneity the rule is `massless'; this is evident in 
the negligible dispersion of the wave packet and its reflection off
the inhomogeneity and then off the left boundary.  With higher 
probability, however, the particle is transmitted through the 
inhomogeneity.  The transmitted wave packet evolves according to a
`massive' rule and begins to disperse slowly so that reflection off 
the right boundary creates a small trailing wave packet.

Recall from Section~5 that the Type~I and Type~II inhomogeneities can
be adjacent with evolution matrix of the form (5.1).  Figure~11 shows
the results of a simulation of this situation:  to the left of the
inhomogeneity $\rho = 0$, $\theta = \pi/4$ and to the right 
$\rho = \pi/4$, $\theta = \pi/3$.  The boundary conditions are of 
Type~III with $\theta' = 0 = \upsilon = \zeta$.  There is the same 
concentration of probability at the inhomogeneity that occurs with the 
Type~I inhomogeneity shown in Figure~9, together with less 
transmission and more dispersion to the right than with the Type~II 
inhomogeneity shown in Figure~10.

\pageinsert
\null\vskip-3\baselineskip
$$
\epsfxsize=\usewidth\epsfbox{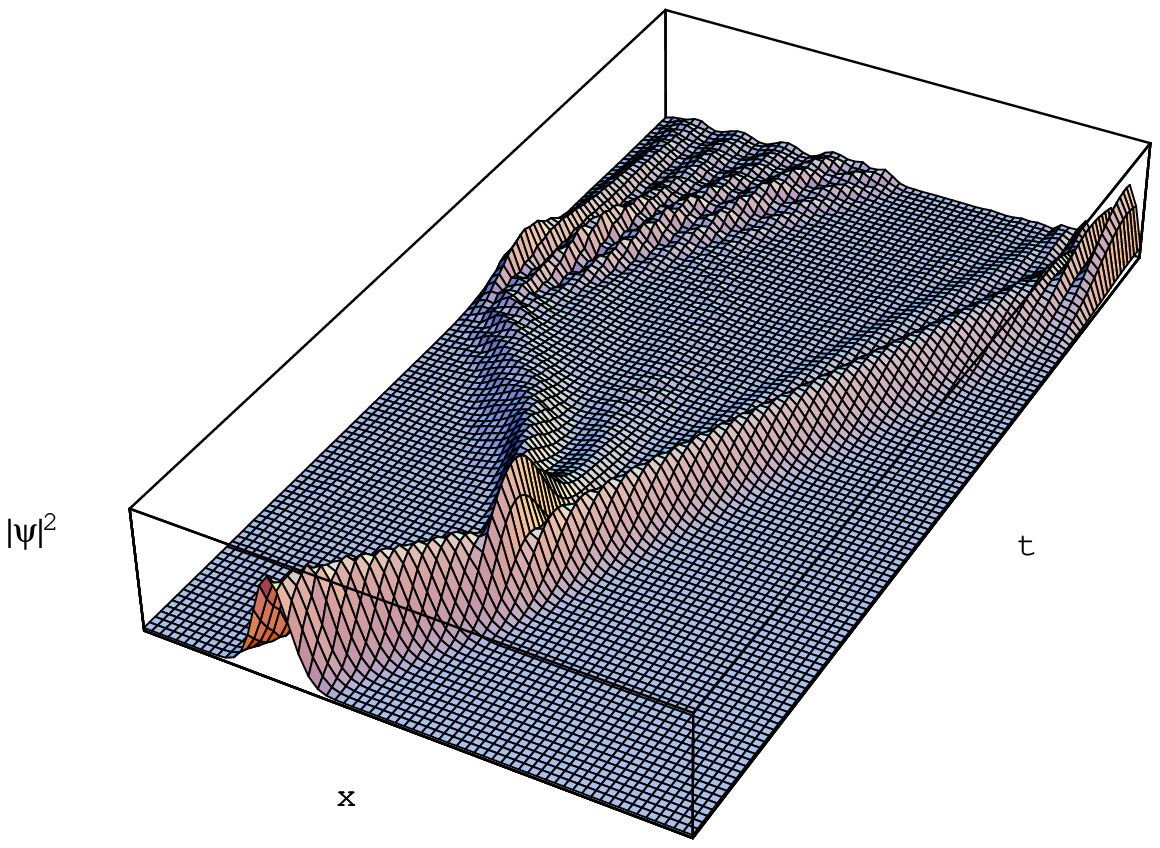}
$$
\vskip-1.5\baselineskip
\eightpoint{%
{\narrower\noindent{\bf Figure~11}.  Evolution of the same wave packet 
as in the previous figures with rule parameters $\rho = 0$, 
$\theta = \pi/4$ to the left and $\rho = \pi/4$, $\theta = \pi/3$ to
the right of a combined Type~I/Type~II inhomogeneity at 
$x = \{31,32\}$.  Both boundaries are of Type~III with 
$\theta = 0 = \upsilon = \zeta$.\par} 
}

\vfill
\null\vskip-3\baselineskip
\vfill
\centerline{({\tenit figure available from author\/})}
\vfill
\vskip-1.5\baselineskip
\eightpoint{%
{\narrower\noindent{\bf Figure~12}.  Evolution of the same wave packet 
as in the previous figures with dual rule parameters $\rho = \pi/3$, 
$\theta = \pi/4$ to the left and $\rho = \pi/4$, $\theta = \pi/3$ to
the right of a combined Type~I/Type~II inhomogeneity at 
$x = \{31,32\}$.  Both boundaries are of Type~III with 
$\theta = 0 = \upsilon = \zeta$.\par} 
}
\endinsert

Finally, recall the $\rho \longleftrightarrow \theta$ duality 
displayed by the dispersion relation (6.2).  Figure~12 shows the 
results of a simulation in the presence of a Type~III inhomogeneity
constructed to convert the rule parameters $\rho = \pi/3$, 
$\theta = \pi/4$ on the left to the dual pair $\rho = \pi/4$, 
$\theta = \pi/3$ on the right.  The reflected and transmitted wave 
packets have more even probabilities than in Figure~11 and evolve with
the opposite group velocities.  There is some asymmetry, most evident 
at the end of the simulation upon reflection from the boundaries; it
is due to the asymmetry of the combined Type~I/Type~II inhomogeneity
as well as of the initial condition.

\medskip
\noindent{\bf 9.  Discussion}
\nobreak

\nobreak
\noindent We have found dual inhomogeneities consistent with unitary 
global evolution of the general one particle rule (1.1)--(1.3):  
Type~I implements a change in $\rho$ while Type~II implements a change
in $\theta$; adjacent Type~I/Type~II inhomogeneities implement changes 
in both $\rho$ and $\theta$.  Each of these three possibilities has a
corresponding boundary condition characterized by additional 
parameters.  Despite this apparent variety of possibilities, the 
unitarity constraint is quite restrictive:  besides the phase 
implementation of an inhomogeneous potential [\BoghosianTaylorSeq,\qmI] 
and some 
degenerate cases, these are the only possible local inhomogeneities up 
to unitary equivalence.

A natural question to ask is how these rule inhomogeneities extend to
the complete multiparticle rule set.  Even the homogeneous rules for 
the general one dimensional QLGA whose one particle subspace we have 
been investigating are quite complicated as there are effectively two,
three and four particle interactions.  In [\qcaqlg], however, we found the 
only particle preserving generalization of the rules (1.1)--(1.3) with 
$\rho = 0$ (\ie, with particles of unit speed).  In this case no more 
than two particles can advect simultaneously to a given lattice site, 
whereupon they scatter in opposite directions with amplitude 
$e^{i\phi}$, $\phi \in \R$.  In fact, the global evolution remains 
unitary if the constant phase angle $\phi$ considered in 
[\DestrideVega,\qcaqlg,\BoghosianTaylorSeq,\lgbu] 
becomes any function of the lattice sites $\phi(x)$.  That is, the two 
particle scattering amplitude can be any inhomogeneous function on the 
lattice, independently of the one particle `scattering' 
amplitudes---and this independent inhomogeneity extends to the values 
of $\phi(x)$ at the boundaries.  (Notice that even in the Type~I and
Type~III boundary conditions where $\overline{w}_0 \not= 0$, and hence
a particle scattering off the boundary can have speed 0, no more than 
two particles can advect to the same lattice site.)  The question of 
determining which boundary conditions are consistent with 
integrability of the model, {\it via\/} the Bethe {\it ansatz\/} 
[\Bethe,\Baxter] as we began studying in [\lgbu] or by generalization of the 
Yang-Baxter equation [\Baxter] as has been used in closely related models 
[\boundaryR], is of fundamental interest.

For the purposes of quantum computation with QLGA, we conclude by 
noting that we have explicitly formulated the possible local 
inhomogeneities in the one dimensional unit speed model.  Extending 
these multiparticle results to multiple speeds and higher dimensions 
seems likely to be algebraically more complicated but conceptually 
similar---single particle single speed models with inhomogeneities 
have been constructed in two dimensions 
[\Riazanov,\VannesteSebbahSornette,\BoghosianTaylorSeq].  More 
interesting is the question of how to exploit such inhomogeneities to 
effect specific quantum computational tasks more efficiently than by 
simply implementing a quantum version of reversible billiard computing 
[\MargolusBB,\Biafore] using a homogeneous rule.  The most natural use 
of QLGA may be to simulate other quantum physical systems; designing 
an inhomogeneous QLGA to be an efficient universal quantum computer 
may consequently be difficult.  A reasonable intermediate goal would 
be to solve specific problems particularly well suited to this 
architecture.  Although neither implements a quantum algorithm, Squier 
and Steiglitz' particle model for parallel arithmetic 
[\SquierSteiglitz] and Benjamin and Johnson's recent proposal for an 
inhomogeneous nanoscale cellular automaton adder [\BenjaminJohnson] 
may provide useful points of departure.

\noindent{\bf Acknowledgements}
\nobreak

\nobreak
\noindent I thank Sun Microsystems for providing support for the 
computational aspects of this project and an anonymous referee for
bringing [\VannesteSebbahSornette] to my attention.

\vfill
\eject

\global\setbox1=\hbox{[00]\enspace}
\parindent=\wd1

\noindent{\bf References}
\bigskip

\parskip=0pt
\item{[\Shor]}
P. W. Shor,
``Algorithms for quantum computation:  discrete logarithms and 
  factoring'',
in S. Goldwasser, ed.,
{\sl Proceedings of the 35th Symposium on Foundations of Computer 
Science}, Santa Fe, NM, 20--22 November 1994
(Los Alamitos, CA:  IEEE Computer Society Press 1994) 124--134.

\item{[\reviews]}
D. P. DiVincenzo,
``Quantum computation'',
\Sc\ {\bf 270} (1995) 255--261;\hfb
I. L. Chuang, R. Laflamme, P. W. Shor and W. H. Zurek,
``Quantum computers, factoring, and decoherence'',
\Sc\ {\bf 270} (1995) 1633--1635;\hfb
A. Barenco and A. Ekert,
``Quantum computation'',
\APS\ {\bf 45} (1995) 1--12;\hfb
J. Preskill,
``Quantum computing:  pro and con'',
preprint (1997) CALT-68-2113, QUIC-97-031, quant-ph/9705032.

\item{[\Schumacher]}
B. Schumacher,
``Quantum coding (information theory)'',
\PRA\ {\bf 51} (1995) 2738--2747.

\item{[\gates]}
A. Barenco, C. H. Bennett, R. Cleve, D. P. DiVincenzo, 
N. Margolus, P. Shor, T. Sleator, J. Smolin and H. Weinfurter,
``Elementary gates for quantum computation'',
\PRA\ {\bf 52} (1995) 3457--3467;\hfb
and references therein.

\item{[\iontraps]}
J. I. Cirac and P. Zoller,
``Quantum computation with cold trapped ions'',
\PRL\ {\bf 74} (1995) 4091--4094;\hfb
C. Monroe, D. M. Meekhof, B. E. King, W. M. Itano and 
  D. J. Wineland,
``Demonstration of a fundamental logic gate'',
\PRL\ {\bf 75} (1995) 4714--4717.

\item{[\cavityQED]}
Q. A. Turchette, C. J. Hood. W. Lange, H. Mabuchi and H. J. Kimble,
``Measurement of conditional phase shifts for quantum logic'',
\PRL\ {\bf 75} (1995) 4710--4713.

\item{[\NMRexp]}
D. G. Cory, M. D. Price, A. F. Fahmy and T. F. Havel,
``Nuclear magnetic resonance spectroscopy:  an experimentally 
  accessible paradigm for quantum computing'',
preprint (1997) quant-ph/9709001;\hfb
I. L. Chuang, N. Gershenfeld, M. G. Kubinec and D. W. Leung,
``Bulk quantum computation with nuclear magnetic resonance:
  theory and experiment'',
preprint (1997).

\item{[\qdots]}
D. L. Loss and D. P. DiVincenzo,
``Quantum computation with quantum dots'',
preprint (1997) cond-mat/9701055.

\item{[\MargolusCAM]}
N. Margolus,
``Ultimate computers'',
in D. H. Bailey, P. E. Bjorstad, J. R. Gilbert, M. V. Mascagni, \etal,
eds., 
{\sl Proceedings of the Seventh SIAM Conference on Parallel Processing
  for Scientific Computing}, San Francisco, CA, 15--17 February 1995
(Philadelphia:  SIAM 1995) 181--186.

\item{[\QCA]}
C. D\"urr, H. L\^e Thanh and M. Santha,
``A decision procedure for well-formed linear quantum cellular 
  automata'',
in C. Puecha and R. Reischuk, eds.,
{\sl STACS 96:  Proceedings of the 13th Annual Symposium on 
     Theoretical Aspects of Computer Science},
Grenoble, France, 22--24 February 1996,
{\sl Lecture notes in computer science} {\bf 1046}
(New York:  Springer-Verlag 1996) 281--292;\hfb
C. D\"urr and M. Santha,
``A decision procedure for unitary linear quantum cellular
  automata'',
in {\sl Proceedings of the 37th Annual Symposium on Foundations 
  of Computer Science},
Burlington, VT, 14--16 October 1996
(Los Alamitos, CA:  IEEE Computer Society Press 1996) 38--45;\hfb
\dajm,
``Unitarity in one dimensional nonlinear quantum cellular automata'',
preprint (1996) quant-ph/9605023;\hfb
W. van Dam,
``A universal quantum cellular automaton'',
preprint (1996).

\item{[\qcaqlg]}
\dajm,
``From quantum cellular automata to quantum lattice gases'',
\JSP\ {\bf 85} (1996) 551--574.

\item{[\Watrous]}
J. Watrous,
``On one-dimensional quantum cellular automata'',
in 
{\sl Proceedings of the 36th Annual Symposium on Foundations of Computer 
  Science}, Milwaukee, WI, 23--25 October 1995
(Los Alamitos, CA:  IEEE Computer Society Press 1995) 528--537.

\item{[\MargolusBB]}
N. Margolus,
``Physics-like models of computation'',
\PD\ {\bf 10} (1984) 81--95.

\item{[\LGAsim]}
T. Toffoli, 
``Cellular automata as an alternative to (rather than an 
  approximation of) differential equations in modeling physics'',
\PD\ {\bf 10} (1984) 117--127;\hfb
\brosl,
``Discrete fluids'',
\LAS\ {\bf 15} (1988) 175--200, 211--217.

\item{[\qmI]}
\dajm,
``Quantum mechanics of lattice gas automata:  One-particle plane waves 
  and potentials'',
\PRE\ {\bf 55} (1997) 5261--5269.

\item{[\BoghosianTaylorsim]}
B. M. Boghosian and W. Taylor, IV,
``Simulating quantum mechanics on a quantum computer'',
preprint (1997) BU-CCS-970103, PUPT-1678, quant-ph/9701019.

\item{[\BoghosianTaylorSeq]}
B. M. Boghosian and W. Taylor, IV,
``A quantum lattice-gas model for the many-particle Schr\"odinger
  equation in $d$ dimensions'',
preprint (1996) BU-CCS-960401, PUPT-1615, quant-ph/9604035, to 
  appear in \PRE.

\item{[\Feynman]}
\feynman,
``Simulating physics with computers'',
\IJTP\ {\bf 21} (1982) 467--488.

\item{[\nanoarch]}
W. D. Hillis,
``New computer architectures and their relationship to physics or
  why computer science is no good'',
\IJTP\ {\bf 21} (1982) 255--262;\hfb
N. Margolus,
``Parallel quantum computation'',
in W. H. Zurek, ed.,
{\sl Complexity, Entropy, and the Physics of Information},
proceedings of the SFI Workshop, Santa Fe, NM, 
29 May--10 June 1989,
{\sl SFI Studies in the Sciences of Complexity} {\bf VIII}
(Redwood City, CA:  Addison-Wesley 1990) 273--287;\hfb
\brosl,
``Parallel billiards and monster systems'',
in N. Metropolis and G.-C. Rota, eds.,
{\sl A New Era in Computation}
(Cambridge:  MIT Press 1993) 53--65;\hfb
R. Mainieri,
``Design constraints for nanometer scale quantum computers'',
preprint (1993) LA-UR 93-4333, cond-mat/9410109.

\item{[\Biafore]}
M. Biafore,
``Cellular automata for nanometer-scale computation'',
\PD\ {\bf 70}\break
(1994) 415--433.

\item{[\lgbu]}
\dajm,
``Quantum lattice gases and their invariants'',
\IJMPC\ {\bf 8} (1997) 717--735.

\item{[\DestrideVega]}
C. Destri and H. J. de Vega,
``Light-cone lattice approach to fermionic theories in 2D.  The
  massive Thirring model'', 
\NPB\ {\bf 290[FS20]} (1987) 363--391.

\item{[\Bethe]}
H. A. Bethe,
``{\it Zur Theorie der Metalle.  I.  Eigenwerte und Eigenfunktionen
       der linearen Atomkette}'',
\ZP\ {\bf 71} (1931) 205--226.

\item{[\Baxter]}
\baxter,
{\sl Exactly Solved Models in Statistical Mechanics\/}
(New York:  Academic Press 1982);\hfb
and references therein.

\item{[\boundaryR]}
I. V. Cherednik,
``Factorizing particles on a half-line and root systems'',
\TMP\ {\bf 61} (1984) 977--983;\hfb
E. K. Sklyanin,
``Boundary conditions for integrable quantum systems'',
\JPA\ {\bf 21} (1988) 2375--2389;\hfb
H. J. de Vega and A. Gonz\'alez Ruiz,
``Boundary $K$-matrices for the six vertex and the 
  $n(2n-1)A_{n-1}$ vertex models'',
\JPA\ {\bf 26} (1993) L519--L524.

\item{[\Riazanov]}
G. V. Riazanov,
``The Feynman path integral for the Dirac equation'',
\SPJETP\ {\bf 6} (1958) 1107--1113.

\item{[\VannesteSebbahSornette]}
C. Vanneste, P. Sebbah and D. Sornette,
``A wave automaton for time-dependent wave propagation in 
  random media'',
\EL\ {\bf 17} (1992) 715--720.

\item{[\SquierSteiglitz]}
R. K. Squier and K. Steiglitz,
``Programmable parallel arithmetic in cellular automata using a
  particle model'',
\CS\ {\bf 8} (1994) 311--323.

\item{[\BenjaminJohnson]}
S. C. Benjamin and N. F. Johnson,
``A possible nanometer-scale computing device based on an adding 
  cellular automaton'',
\APL\ {\bf 70} (1997) 2321--2323.

\bye